\documentclass{ws-ijmpd}

\begin{document}

\markboth{Mark J. Reid}
{}

%
\catchline{}{}{}{}{}
%

\title {IS THERE A SUPERMASSIVE BLACK HOLE\\
AT THE CENTER OF THE MILKY WAY?}


\author{Mark J. Reid}

\address{Harvard--Smithsonian Center for Astrophysics\\
         60 Garden Street, Cambridge, MA 02138, USA\\
         mreid@cfa.harvard.edu}


\def\Rsch           {\ifmmode {R_{\rm Sch}} \else  {$R_{\rm Sch}$} \fi}
\def\MoverR         {\ifmmode {M/R} \else  {$M/R$} \fi}
\def\Mh             {\ifmmode {M_h} \else  {$M_h$} \fi}
\def\rhoh           {\ifmmode {\rho_h} \else  {$rho_h$} \fi}
\def\Msun           {M$_\odot$}
\def\Msunpercpc     {M$_\odot$~pc$^{-3}$}
\def\kms            {km~s$^{-1}$}
\def\etal           {et al.}
\def\ie             {i.e.}
\def\eg             {e.g.}
\def\NeII           {Ne~{\small II}}
\def\um             {$\mu$m}
\def\SgrA           {Sgr~A*}
\def\SgrAwest       {Sgr~A-West}
\def\SgrAeast       {Sgr~A-East}
\def\p              {\phantom{>}}
\def\q              {\phantom{0}}

\maketitle

\begin{history}
\received{2008 August 18}
\revised{}
\comby{Managing Editor}
\end{history}

\begin{abstract}
This review outlines the observations that now 
provide an overwhelming scientific case that the center of our
Milky Way Galaxy harbors a supermassive black hole. 
Observations at infrared wavelength trace stars that 
orbit about a common focal position and require a central mass ($M$) 
of $4\times10^6$~\Msun\ within a radius of 100~AU.  Orbital 
speeds have been observed to exceed 5,000~\kms.  At the
focal position there is an extremely compact radio source (\SgrA),
whose apparent size is near the Schwarzschild radius ($2GM/c^2$).
This radio source is motionless at the $\sim1$~\kms\ level at the 
dynamical center of the Galaxy.  The mass density required by these 
observations is now approaching the ultimate limit of a supermassive 
black hole within the last stable orbit for matter near the event horizon.  
 
\end{abstract}


\vskip 0.4truecm
\centerline {\bf Table of Contents}
\vskip 0.2truecm
1. Introduction  \hfill{2.}
\vskip 0.0truecm
2. Evidence for Supermassive Black Holes (SMBHs) \hfill{2.}
\vskip 0.0truecm
3. Our Galactic Center \hfill{5.}  
\vskip 0.0truecm
~~~~~3.1  The Discovery of \SgrA\  \hfill{5.}
\vskip 0.0truecm
~~~~~3.2  The Growing Case for a Dark Mass at the Galactic Center   \hfill{7.}
\vskip 0.0truecm  
4. Recent Overwhelming Evidence that \SgrA\ is a SMBH \hfill{8.}
\vskip 0.0truecm
~~~~~4.1  Stars Orbiting an Immense Unseen Mass Concentration  \hfill{9.}
\vskip 0.0truecm
~~~~~4.2  The Unseen Mass is Centered on \SgrA\  \hfill{10.}
\vskip 0.0truecm
~~~~~4.3  \SgrA's Emission is Extremely Compact  \hfill{13.}
\vskip 0.0truecm
~~~~~4.4  \SgrA\ is Motionless at Dynamical Center of Milky Way   \hfill{14.}
\vskip 0.0truecm
5. Excluding Alternatives to a SMBH \hfill{17.}
\vskip 0.0truecm
6. Other Evidence for a SMBH \hfill{18.}
\vskip 0.0truecm
7. Broader Implications of SMBHs at Centers of Galaxies  \hfill{19.} 
\vskip 0.0truecm
8. Future Observations \hfill{19.} 
\vskip 0.0truecm
9. Summary  \hfill{20.} 
\newpage

\section {Introduction}

In the late 18$^{th}$ century, the English naturalist John Michell and
the French mathematician Pierre Simon Laplace considered what would 
happen if a huge mass were placed in an incredibly small volume.
They conjectured that gravitational forces might
not allow anything, even light, to escape.  Two centuries later,
Albert Einstein's theory of General Relativity provided the theoretical 
foundation for such conjectures, and in the 1960s John Archibald 
Wheeler introduced the term ``black hole'' to describe the effects of 
mass at such an extreme density.

        The concept of a black hole formed by the explosive
collapse of a dying star is astounding.  The possibility that matter 
from millions and even billions of stars can condense into a single 
supermassive black hole (SMBH) is even more fantastic.  
Yet we are now confident that supermassive black holes exist and in 
fact are commonplace, occupying the centers of many, if not all,
of the $\sim10^{11}$ galaxies in the Universe.    
Indeed, SMBHs may hold more than 0.01\% of the baryonic mass of the 
Universe.

\section   {Evidence for Supermassive Black Holes} \label{section:evidence}

\begin{figure*}
\begin{center}
\begin{tabular}{cc}
\includegraphics[width=0.97\textwidth]{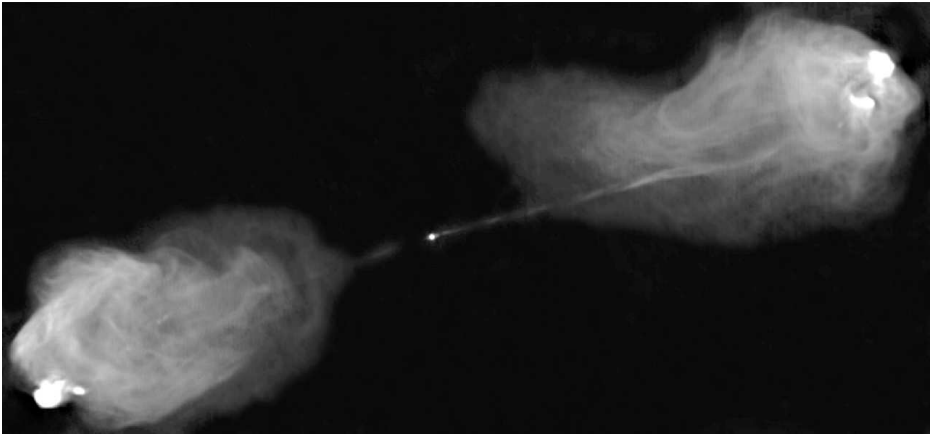} \\
\end{tabular}
\end{center}
\caption{Image of the radio galaxy Cygnus A made with the
Very Large Array.  The two large lobes of synchrotron emitting
plasma dominate.  In some radio galaxies these lobes can span
$\sim10^6$~parsecs (1 pc $\approx3\times10^{16}$~m $\approx 3.3$ 
light years), and the minimum energy to power the lobes can be up to 
$10^7~{\rm M}_\odot c^2$~!  Note the thin ``jets'' that connect 
the small bright nucleus at the center with the lobes and terminate with
``hot spots'' at the outer edges of the lobes.  An optical image of the 
central galaxy would be comparable in size to a hot spot.
All of the energy to power the lobes comes from the nucleus, in fact 
from a region smaller than 1~parsec, the distance from the Sun to the 
nearest star.  Image courtesy of NRAO/AUI. 
        }
\label{fig:cyga}
\end{figure*}

Early evidence for SMBHs closely paralleled the development
of radio astronomy.  Very strong sources of radio waves were 
discovered in the early years of radio astronomy.
Accurate positions of these sources revealed that they were often
centered on distant galaxies.  In the 1950s, radio interferometers
revealed a totally unexpected picture of these ``radio galaxies.''  
The radio waves did not come from the galaxy itself, 
but from two giant ``lobes'' symmetrically placed about, 
but well separated from, the parent galaxy (see Fig.~\ref{fig:cyga}).  
These lobes can be among the largest structures in the Universe, 
hundreds of times the size of the parent galaxy.

	How are immense radio lobes energized?  Their symmetrical
placement about galaxies indicated a galactic link.  In the 1960s,
sensitive radio interferometers confirmed the previously 
circumstantial case by detecting faint trails (called ``jets'') of 
radio emission from the lobes back toward the parent galaxy.  
Radio jets often lead back to a point-like source at the precise 
center of the galaxy.  These point-like sources were found to
be variable on timescales less than one year, implying sizes
less than a light-year\cite{Dent:65,Maltby:65}.  
Martin Rees, in a remarkably prescient paper, showed that the
source energetics and variability required synchrotron emitting 
plasma that has been accelerated to bulk relativistic speeds\cite{Rees:66}.  
One {\it predicted} consequence was that such
sources would display apparent motions on the sky that were
faster than the speed of light (so called super-luminal motion discussed
below).

      In the late 1960s, Very Long Baseline Interferometry (VLBI) 
extended baselines to the size of the Earth, achieving
resolution ($\approx\lambda/D$, where $\lambda$ is the observing 
wavelength and $D$ is the maximum separation of interferometer elements) 
better than 0.001~arcseconds\cite{Cohen:68}.  
This was achieved by replacing
cables between antennas of the interferometer with tape recorders
synchronized with atomic oscillators.
Radio images made from VLBI observations revealed that the 
sources at the centers of radio galaxies are incredibly small,
even smaller than the distance between the Sun and the nearest 
star\cite{Kellermann:68}.

Simple calculations of the {\it minimum} energy needed to power 
giant radio lobes require the {\it total} 
conversion of up to $10^7$ stars into energy\cite{Kellermann:88}!  
Since nuclear reactions convert less than 1\% of mass 
to energy, trying to explain a radio galaxy with nuclear power would 
require channeling more than $10^9$ stars through a region smaller 
than the distance between the Sun and the nearest star.  
Because of these requirements, astronomers began considering a more 
efficient energy source: a supermassive black hole.

The small measured sizes of radio sources at the centers of galaxies
were consistent with those inferred 
from the rapid variability.  Additionally, VLBI data taken months
to years apart indicated that ``blobs'' of synchrotron emitting
plasma appeared to move across the sky at speeds exceeding that
of light\cite{Whitney:71,Cohen:71}.  
This phenomenon, called super-luminal motion, has now 
been well documented in numerous sources\cite{Zensus:97}.  
Super-luminal motion is simply explained as an ``optical illusion'' that 
occurs when light emitting plasma moves toward us, nearly along our
line of sight, at speeds near, but below, that of light. 
This phenomenon finds a natural explanation in the acceleration of 
material from regions very close to the event horizon of a black 
hole\cite{Blandford:77}. 

In 1990s, images made with the Hubble Space Telescope\cite{Ford:94,Harms:94}
showed that material located within 18 parsecs\footnote{1 parsec (pc) = 
3.26 light-years, approximately the distance from the Sun to the nearest 
stars.} of the center of the galaxy M~87 was moving with speeds of about 
750 \kms.  Assuming the motions are from material 
in orbit about a large mass concentration, the Virial theorem implies a 
central mass of $\approx2.4\times10^9$~\Msun\ (1 solar mass or \Msun
$=2\times10^{30}$ kg) and an average mass density
of $10^5$~\Msunpercpc.  While this density is very high, 
it is not above stellar densities in dense clusters and 
does not require a supermassive black hole.  

\begin{figure*}
\begin{center}
\begin{tabular}{cc}
 \includegraphics[width=0.90\textwidth]{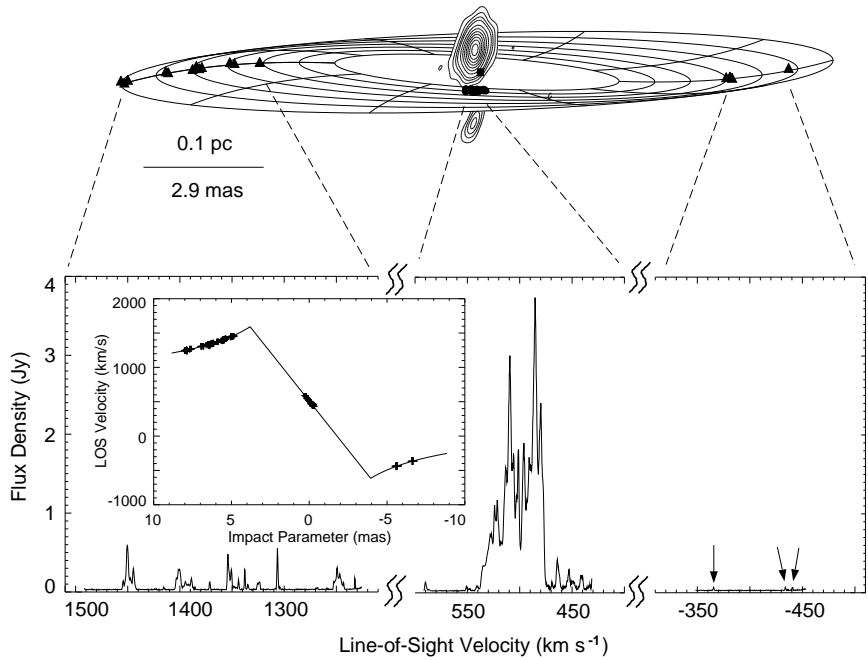}
\\
\end{tabular}
\end{center}
\caption{Water vapor masers in a thin rotating disk at the center
of the galaxy NGC~4258 from Herrnstein \etal\ (1999). 
{\it Top:} Schematic drawing of the disk superposed with a radio frequency 
image ({\it contours}) of synchrotron emitting plasma jets emanating from 
the center.  Water masers are shown as black triangles in the disk.  
{\it Bottom:} An observed spectrum (flux density versus Doppler velocity).
{\it Inset:} Plot of maser Doppler velocity versus distance from the center.
A Keplerian ($v \propto 1/\sqrt{r}$) rotation curve is shown with
the curved solid lines.  The straight line shows the changing line of sight 
projection of masers from the center of the spectrum that lie on a 
fixed radius in a small arc in front of the disk center.
(Reprinted by permission of Macmillan Publishers Ltd: Nature, {\bf 400}, 
1999, 539)
        }
\label{fig:ngc4258}
\end{figure*}

In the 1980s, maser emission was discovered toward the centers of nearby 
galaxies\cite{Claussen:84}.  This emission comes from clouds of gas
containing trace amounts of water molecules whose level populations for 
the $6_{16}-5_{23}$ transition at 22 GHz become inverted.  In the mid-1990s, 
water masers in the galaxy NGC~4258 were discovered to have internal motions 
exceeding 1000 \kms~\cite{Nakai:93}.  Subsequent VLBI images of the 
masers showed that they came from a rotating disk of 
material\cite{Miyoshi:95,Greenhill:95,Herrnstein:99} 
(see Fig.~\ref{fig:ngc4258}).  
The variation of velocity ($v$) with
radius ($r$) followed $v\propto1/\sqrt{r}$ to better than 1\% accuracy,
indicating gravitational orbits about a compact central mass. 
The rotation speed was about 1000 \kms\ at a radius
of about 0.13 parsecs, requiring a central gravitational mass of 
$4\times10^7$~\Msun.   The corresponding enclosed mass density is  
$4\times10^9$~\Msunpercpc.  Were one to place $4\times10^7$ stars inside a 
radius of 0.13 parsecs, the system would be dynamically unstable with less massive
stars being expelled (``evaporated'') and more massive stars sinking to the 
center, colliding and possibly forming a black hole.  The timescale for the 
cluster to evaporate would be fairly short, $\sim10^9$~years,
making it unlikely that a cluster of stars could provide the central 
gravitational mass\cite{Maoz:98}.  However, the densities were not extreme 
enough to conclusively rule out clusters of some types of objects or more 
exotic speculations involving dense condensations of elementary particles.  

\begin{figure*}
\begin{center}
\begin{tabular}{cc}
\includegraphics[width=0.95\textwidth]{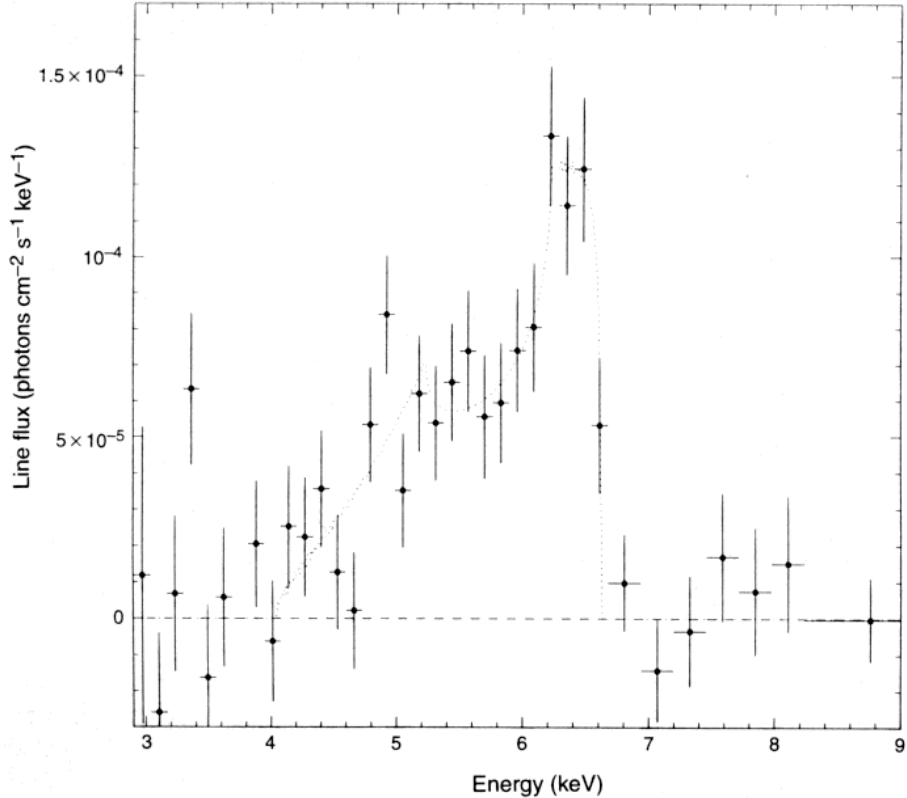} \\
\end{tabular}
\end{center}
\caption{Fluorescence Fe K$\alpha$ line toward the galaxy 
MCG-6-30-15 from Tanaka \etal\ (1995).  The line is extremely
broad ($\sim10^5$ \kms), asymmetric, and mostly redshifted
from the rest energy of about 6.4 keV.  This line probably 
arises form material in an accretion disk inside of $10 \Rsch$
of a SMBH where Doppler and gravitational redshifts are large.
(Reprinted by permission of Macmillan Publishers Ltd: Nature, 
{\bf 375}, 1995, 659)
        }
\label{fig:mcg-6-30-15}
\end{figure*}

In 1994, the ASCA satellite's x-ray telescope recorded a spectrum of
the nucleus of the galaxy MCG-6-30-15, showing a broad line at 6.4 keV
from iron K$\alpha$ emission\cite{Tanaka:95}.  
As shown in Fig.~\ref{fig:mcg-6-30-15}, the full width at zero intensity 
was $\sim10^5$~\kms, or about 30\% of the speed of light!  
The line is asymmetric with most of the 
flux appearing redshifted with respect to the rest motion of the galaxy.
The simplest interpretation of this line is that it comes from
fluorescence of iron atoms in an accretion disk irradiated by a halo
of hot gas and all within $\sim10\Rsch$ of a massive black hole,
where $\Rsch=2GM/c^2$ is the Schwarzschild radius. 
Thus, by the mid-1990s extremely strong, but perhaps not overwhelming, 
evidence for supermassive black holes existed.

\section {Our Galactic Center}

\begin{figure*}
\begin{center}
\begin{tabular}{cc}
 \includegraphics[width=0.47\textwidth]{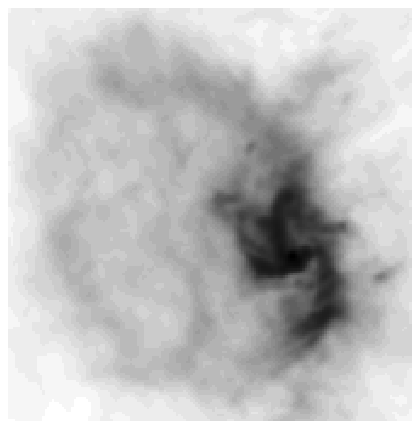} &
 \includegraphics[width=0.47\textwidth]{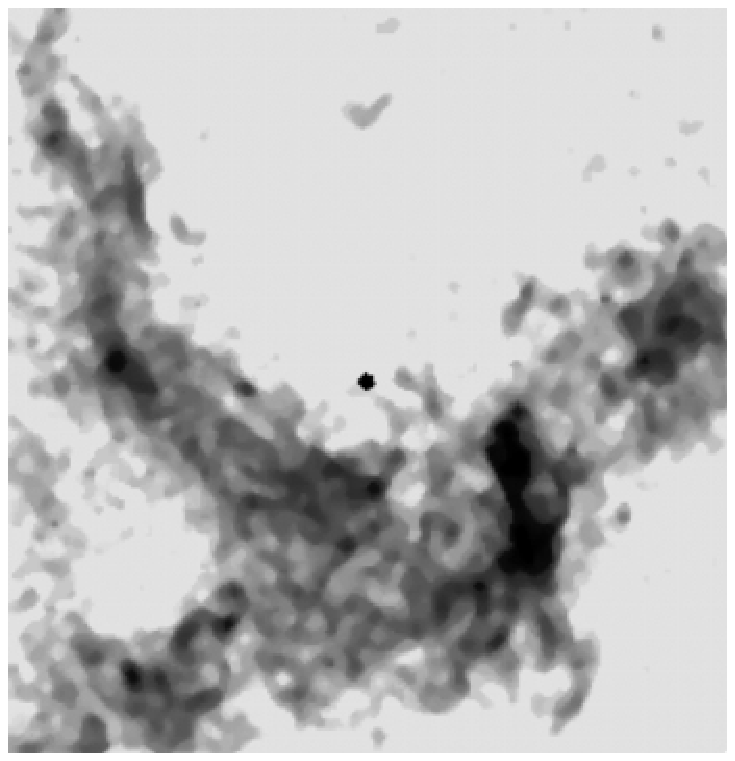}
\\
\end{tabular}
\end{center}
\caption{Radio wavelength negative images of the center  
of the Milky Way with East toward the left and North toward the top.
{\it Left Panel:} The multi-component source Sgr A.  The diffuse
elliptical shell-like source \SgrAeast, a supernova remnant from an
exploded star, fills most of the left side of this panel, which measures
$\approx7$ parsecs across.  The bright spiral-shaped emission 
toward the right-center of the panel is called \SgrAwest\ and comes from 
plasma spiraling inward to the center.
{\it Right Panel:} Expanded view of the central 0.25 parsecs, showing the
central region of the \SgrAwest\ spiral.  Also seen at the center of 
the panel is a point-like source called \SgrA, a candidate for a 
supermassive black hole.  (Images courtesy of J.-H. Zhao.)
        }
\label{fig:sgra}
\end{figure*}

     Surveys of the sky at radio wavelengths in the 1950s revealed
a strong radio source in the constellation Sagittarius toward
the center of the Milky Way.  This radio source was
named Sagittarius~A (Sgr~A), where the letter ``A'' denoted the strongest
source in the constellation.  Early radio telescopes did not have
the angular resolution to resolve this source, and its nature
remained a mystery for some time.  With the advent of radio
interferometry, Sgr~A was revealed to contain multiple 
components\cite{Morris:96}.
Fig.~\ref{fig:sgra} shows recent images of this complicated region made 
the Very Large Array radio interferometer.  The left-hand image reveals 
at least two strong and extended sources: \SgrAeast, a non-thermal 
(synchrotron emitting) supernova remnant and \SgrAwest, 
a ``spiral-shaped'' source of thermal bremsstrahlung emission
from ionized gas. 

\subsection {The Discovery of \SgrA} \label{section:sgra_discovery}

In 1974, a very compact radio source, smaller than 1 arcsecond in 
diameter, was discovered toward Sagittarius~A
and later given the name Sagittarius~A* (\SgrA)\cite{Balick:74}. 
Early VLBI observations established that \SgrA\  was extremely compact -- 
less than the size of our solar system when observed at centimeter 
wavelengths\cite{Lo:85}.  However, no obvious optical, infrared, or 
X-ray counterpart to \SgrA\  could be easily identified, and its nature 
remained a mystery.   

Interestingly, even before the discovery of a compact radio source 
at the center of the Milky Way, 
it had been speculated that the center of our Galaxy contained a 
supermassive black hole\cite{Lynden-Bell:71}.  This speculation
was based primarily on two arguments: 1) the presumed similarity 
between the nuclei of radio (and other active) galaxies and that of our
Galactic center and 2) the possibility of explaining most of the 
luminosity arising from the Galactic center as owing ultimately
to accretion of material onto a black hole.  While argument 1) has
been borne out, argument 2) has not, since most of the luminosity
of the Galactic center can be traced to stellar origins.  
Unlike the centers of active galaxies believed
powered by supermassive black holes, \SgrA\ is extremely 
under-luminous\cite{Ozernoy:96}.

\subsection {The Growing Case for a Dark Mass at the Galactic Center} 

     In the late 1970s, infrared observations of gas motions provided the 
first clues for a very large mass concentration at the center of the 
Milky Way.  Analysis of the center velocities of fine-structure lines of
singly-ionized neon atoms (at a wavelength of 12.8~\um) across \SgrAwest\  
showed differences of $\pm260$~\kms.  
Assuming that the gas clouds were in circular orbits indicated ``a central 
point-like mass of several $\times 10^6$~\Msun\ in addition to several 
$\times 10^6$~\Msun\ of stars within 1~parsec of the center''\cite{Lacy:80}.
This proved to be a remarkably accurate inference.
While these observations provided the first solid observational evidence for
a supermassive ``object'' at the center of the Milky Way, other possibilities 
were recognized.  Firstly, while a point-mass component fit the data best, 
a distributed mass of $\sim 10^7$~stars could not be ruled out.  Secondly,
gas is susceptible to non-gravitational forces, and the assumption of
gravitationally bound orbits was questioned.
Finally, the ``point-mass'' component needed only to be smaller than
$\approx0.2$~parsecs, and as discussed later in \S\ref{section:alternatives}, 
this does not require a SMBH.
  
While gas is susceptible to non-gravitational forces, stars are not, and soon
measurements of radial velocities (Doppler shifts) of stars confirmed the 
presence of several 
million solar masses located within $\sim0.1$~parsecs of the center of the Milky 
Way\cite{McGinn:89,Sellgren:90,Krabbe:95,Haller:96,Genzel:97}.  
There is indeed a dense cluster of stars at the center of the Milky Way,
which cannot be seen at optical wavelengths because visible light is totally 
absorbed by dust between the Galactic center and the Sun.
However, these stars can be seen at infrared wavelengths, where $\sim10$\%
of the 2~\um\ wavelength light is received.  
Using novel techniques that allow diffraction-limited imaging in the 
infrared, groups led by Reinhard Genzel in Germany and Andrea Ghez in the 
USA have been measuring positions of these stars for more than a decade
\cite{Eckart:96,Eckart:97,Ghez:98,Genzel:00}.  These results
showed that stars projected very close to the position of \SgrA\  
were moving very rapidly across the sky.  

\begin{figure*}
\begin{center}
\begin{tabular}{cc}
\includegraphics[width=0.49\textwidth]{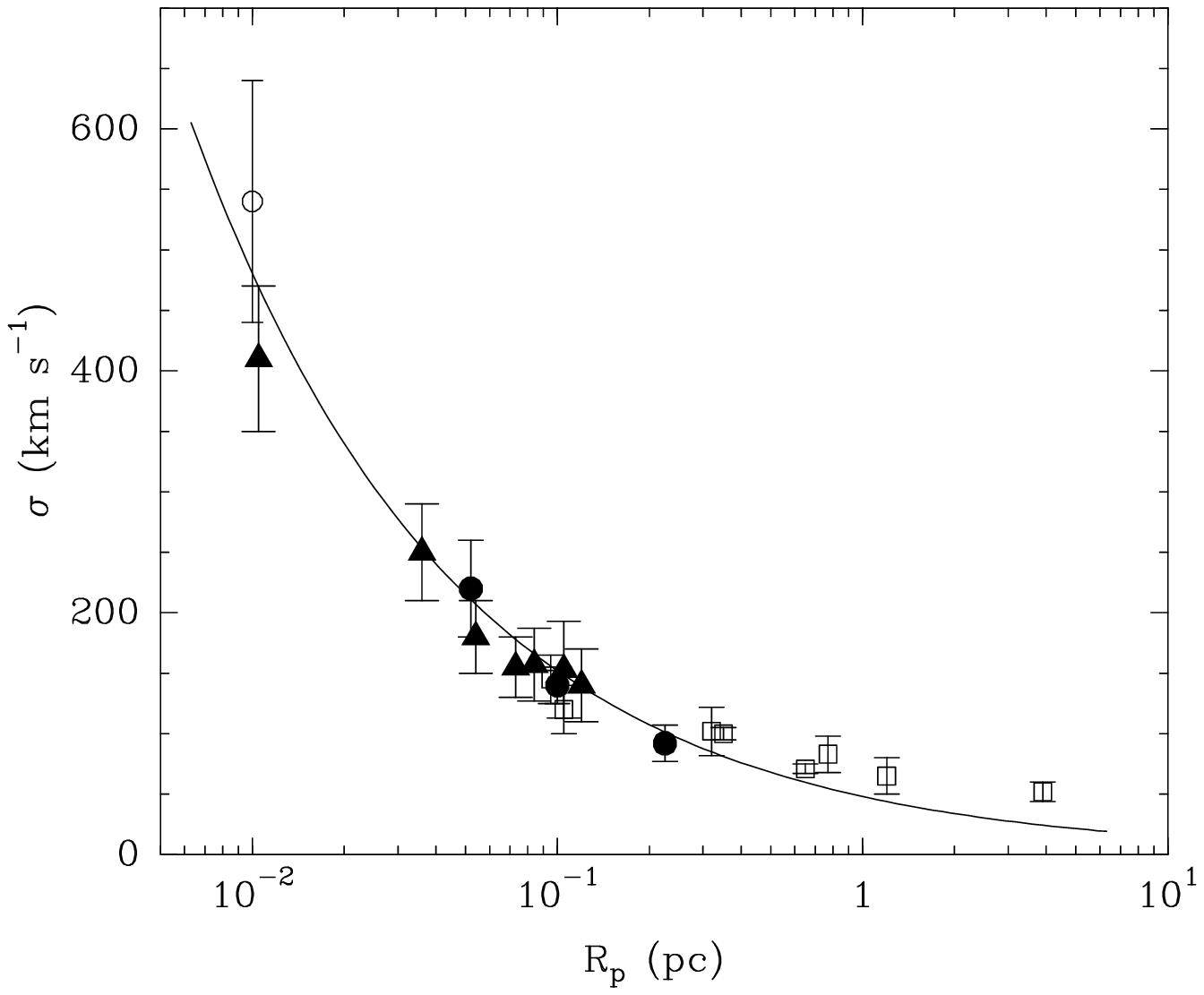} & 
\includegraphics[width=0.47\textwidth]{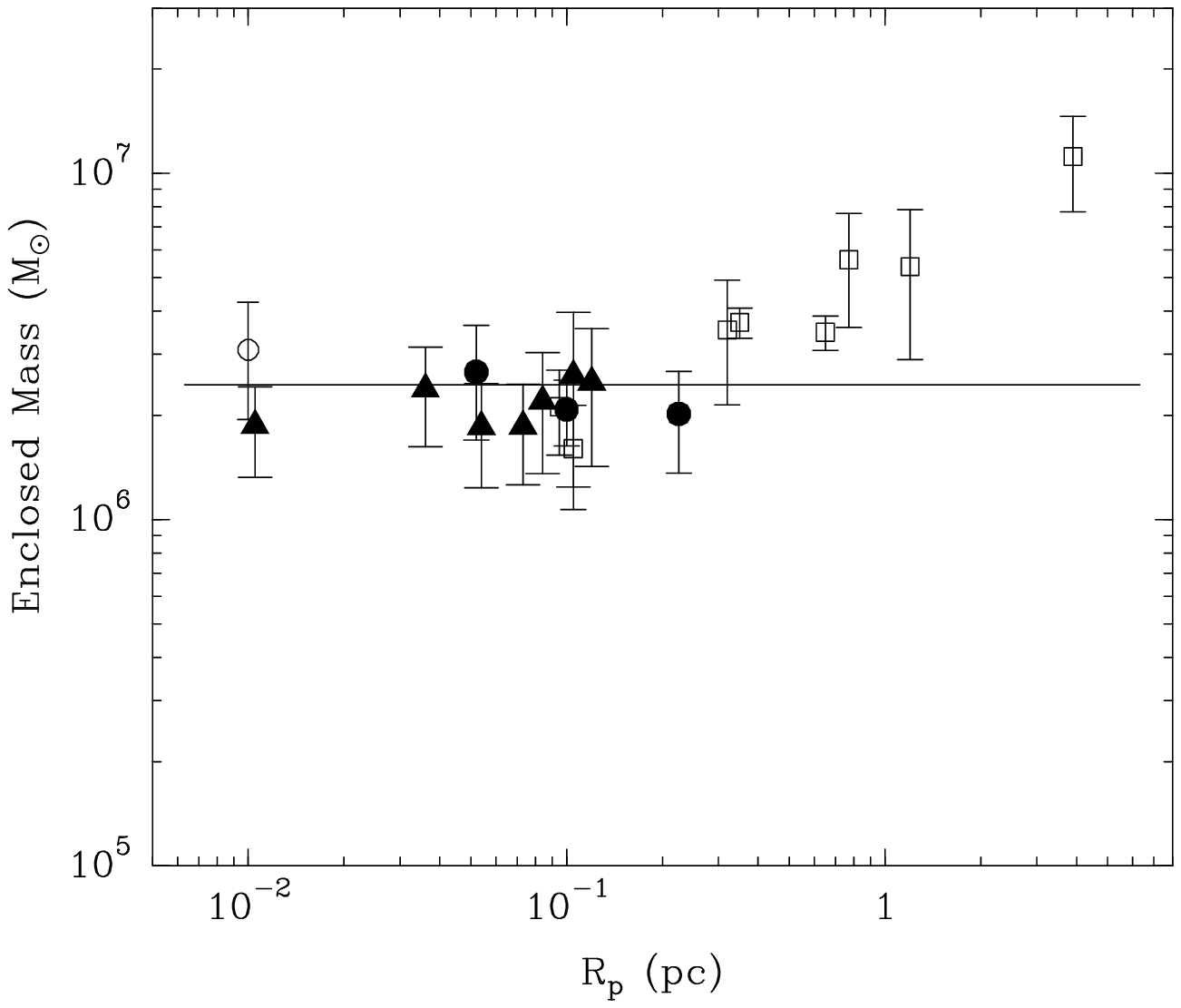} 
\\
\end{tabular}
\end{center}
\caption{Measured stellar velocity dispersions {\it (left panel)} and 
inferred enclosed mass {\it (right panel)} versus projected distance 
from the Galactic center (after Eckart \& Genzel 1997 and Ghez et al 1998).  
The line through the data corresponds to Keplerian orbits for a point-like 
mass of $2.5\times10^6$~\Msun\ and fits the data well for radii smaller than
0.2~pc.  Beyond this radius, the velocity dispersions exceed that
expected from a point mass, owing to the contribution to the enclosed mass
from an observed dense cluster of stars.
        }
\label{fig:enclosed_mass}
\end{figure*}

Fig.~\ref{fig:enclosed_mass} plots the stellar velocity dispersions available 
in 1998 and the inferred enclosed mass as a function of projected distance 
from the position of \SgrA.  Measurements of stellar radial velocity sample 
projected distances from 4 to 0.1 parsecs and proper motions
(motions on the plane of the sky) sample from 0.3 to 0.01 parsecs.
Between projected radii of 0.2 and 0.01 parsecs, the velocity dispersion,
$\sigma$, increases as $\sigma \propto 1/\sqrt{r}$.  
The enclosed mass estimated for virialized material is indeed nearly 
constant between these radii, greatly strengthening the case for a large 
``point mass.''  The implied mass density, while very high (comparable to 
that inferred from the water masers in the galaxy NGC~4258 discussed in
\S\ref{section:evidence}), still could not rule some alternatives to a SMBH 
(see \S\ref{section:alternatives}).

\section  {Recent Overwhelming Evidence that \SgrA\  is a SMBH}

Evidence for the existence of SMBHs, especially for \SgrA, has been 
steadily growing over the years, but recently observational constraints 
have become so strong that there can be almost no doubt that \SgrA\ is a 
supermassive black hole.  

\subsection  {Stars Orbiting an Immense Unseen Mass Concentration}

\begin{figure*}
\begin{center}
\begin{tabular}{cc}
\includegraphics[width=0.95\textwidth]{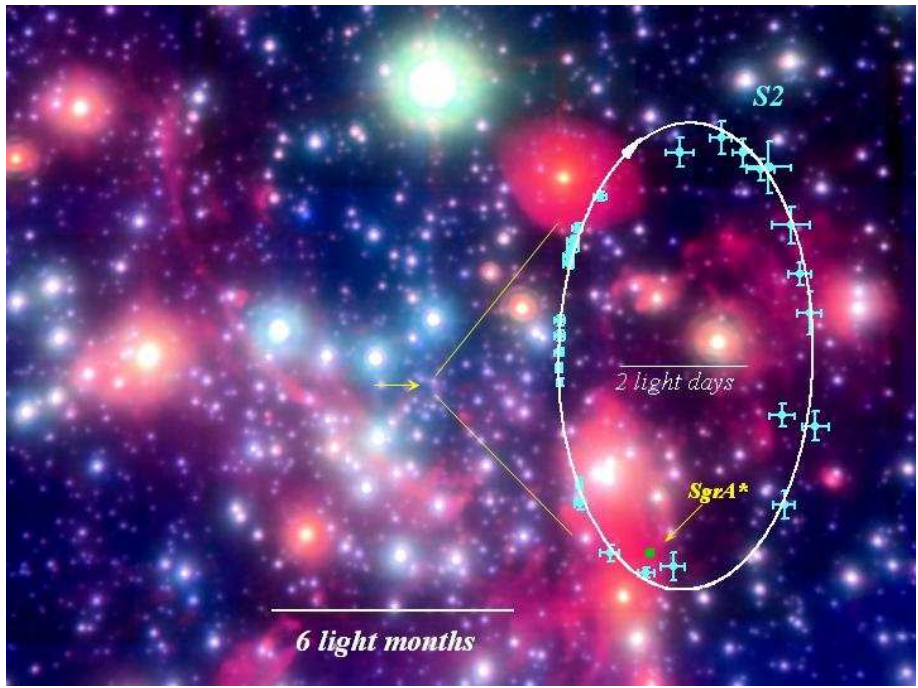} 
\end{tabular}
\end{center}
\caption{False color infrared image taken with by the European Southern
Observatory's Very Large Telescope of the central few parsecs of the Milky Way.  
Superposed with a 100-times finer scale
is the orbital track of one star named S2.  The orbital period of 
S2 is 15.8 years, and recently a complete and closed elliptical orbit 
has been observed.  The orbit requires an unseen mass of 
$\approx4\times 10^6$~\Msun\ at the focal position, 
indicated by the arrow.  The focal position is coincident with the position 
of the compact radio source \SgrA\ as discussed in 
\S\ref{section:sgra_position}.
(Image courtesy R. Genzel.)
}
\label{fig:S2_orbit}
\end{figure*}

\begin{figure*}
\begin{center}
\begin{tabular}{cc}
\includegraphics[width=0.95\textwidth]{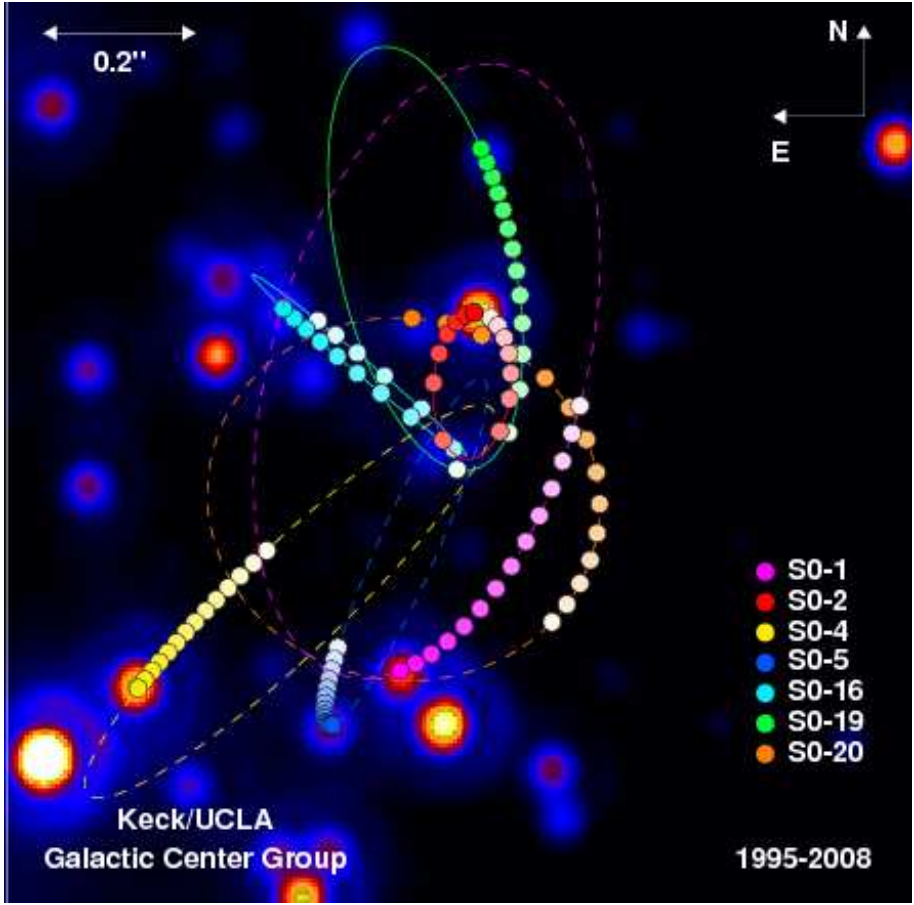} \\
\end{tabular}
\end{center}
\caption{
Stars within the 0.02 parsecs of the Galactic center orbiting an unseen mass.
Yearly positions of seven stars are indicated with filled colored circles.
Both curved paths and accelerations (note the non-uniform spacings between 
yearly points) are evident. Partial and complete elliptical orbital fits for 
these stars are indicated with lines.   All orbital fits require 
the same central mass of $\approx4\times10^6$~\Msun\ and a common 
focus at the center of the image, the position of the radio source \SgrA.  
(Image courtesy A. Ghez.)
        }
\label{fig:stellar_orbits}
\end{figure*}

Continued monitoring of the positions of stars, with increasing
positional accuracy, led to clear detections of acceleration
(ie, curving motions on the sky)\cite{Ghez:00,Eckart:02}.  Importantly, the
directions of the acceleration vectors ``pointed'' to a common central
gravitational source very close to the position of \SgrA.  
Recently these observations culminated in the discovery that
stars are executing elliptical paths (orbits) on the sky
\cite{Ghez:00,Schoedel:02,Ghez:03,Schoedel:03,Ghez:05}.  
One star named S2 (a.k.a. S0-2) and has now been observed over one 
complete 15.8-year elliptical orbit (see Fig.~\ref{fig:S2_orbit}).
All stellar orbits are well fit by a single enclosed mass and  
focal position (see Fig.~\ref{fig:stellar_orbits}).  
Two stars have been observed to approach within 100~AU
\footnote{1 Astronomical Unit (AU), the mean distance from the Earth 
to the Sun, equals $1.5\times10^{11}$~m or $5\times10^{-6}$ parsecs.}
of the focal position, moving at nearly $10^4$ \kms!  
{\it The orbital solutions leave no doubt that the stars are responding 
to an unseen compact mass of $\approx4\times10^6$~\Msun.}  

   We note that the central mass estimated from stellar velocity 
dispersions ($\approx2.5\times10^6$~\Msun) has been lower than the 
value obtained from fitting Keplerian orbits ($\approx4\times10^6$~\Msun).
Transforming one-dimensional velocity dispersions to enclosed masses can 
be accomplished with a standard Virial analysis or using alternative 
statistical mass estimators: \eg\ Bahcall-Tremaine\cite{Bahcall:81}.  
These methods are usually based on an assumption of isotropic motions.  
While this is well justified for old stars orbiting the Galactic center, 
it does not work well for young stars that still ``remember'' the orbital 
plane of the gas cloud from which they recently formed.  At least one such 
collection of young stars has been discovered\cite{Genzel:00,Lu:08}.
An alternative statistical mass estimator, the 
Leonard--Merritt\cite{Leonard:89} approach, does not depend on
isotropic distributions of stellar velocities.  However, as for all 
projected mass estimators, it assumes measurements over all radii.  
For the central stellar cluster, which is distributed as a power-law 
over the range of observed radii\cite{Genzel:03b,Schoedel:07}, 
normalization problems arise.  
Empirical corrections to the Leonard--Merritt estimator, for the limited 
range of projected orbital radii of stars that have been measured, 
increase the central mass estimate and resolve the discrepancy with 
orbital fitting\cite{Genzel:00}.

The possibility of a combination of a point mass and an extended ($>100$ AU) 
distribution of mass has been considered\cite{Mouawad:05,Ghez:08,Gillessen:08}.
Based on the small deviations from an elliptical orbit for star S2 allowed
by measurement uncertainty, any extended component within 0.01 parsecs of
the center must be $<10$\% of the point mass.  Limits on deviations of orbits
from those responding to a pure point-mass will undoubtedly improve rapidly
as the star S2 proceeds on a second cycle and other stellar orbits are
better traced.  Because the orbital paths are almost perfect ellipses,
most of the unseen mass must be contained within a radius of about 
100~AU (0.0005 pc).  This implies a mass density of 
$>8\times10^{15}$~\Msunpercpc, which is so great that one can rule out 
the exotic speculation that a ``ball'' of Fermions with rest energies 
of $\sim15$~keV might provide the extreme central mass for \SgrA\ and other, 
more massive, galactic nuclei (see \S\ref{section:alternatives}) .  
sss
\subsection {The Unseen Mass is Centered on \SgrA} \label{section:sgra_position}

     The infrared results just described are beautifully complemented
by observations at radio wavelengths.  It is crucial to know the position
of \SgrA\ on infrared images.  However, while \SgrA\ is a strong radio
source, it is extremely dim at infrared wavelengths.  In general, the infrared
coordinate system is not known to better than about 0.1 arcseconds relative
to the International Celestial Reference Frame determined by VLBI observations
at radio wavelengths.  Unfortunately, there are many stars in the 
Galaxy's central 0.1 arcseconds, so this level of position accuracy
is inadequate to locate and identify \SgrA\ on crowded infrared images.

\begin{figure*}
\begin{center}
\begin{tabular}{cc}
\includegraphics[width=0.90\textwidth]{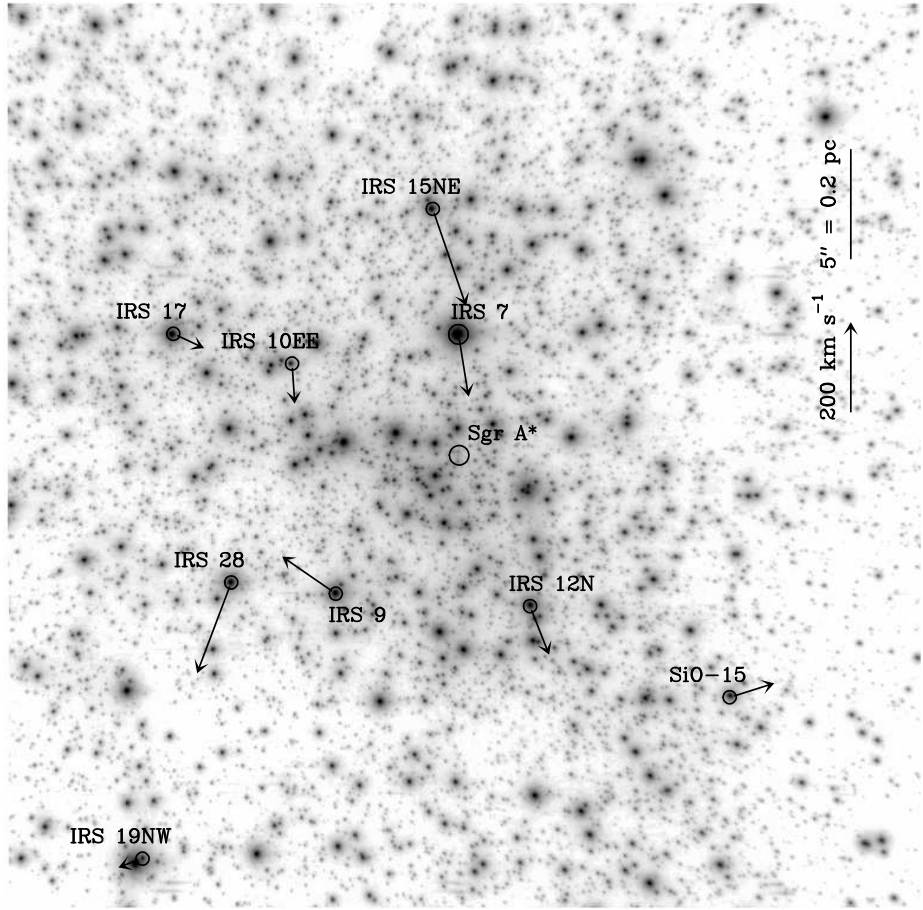}  \\
\end{tabular}
\end{center}
\caption{Infrared negative image at  2.2~\um\ wavelength of the 
central $\pm20$ arcseconds of the Milky Way from Reid et al. (2007).
The positions ({\it circles}) and measured motions ({\it arrows}) of red
giant stars that have detected radio emission from circumstellar 
SiO masers are indicated.  These stars were used to accurately transfer 
the position of \SgrA, measured in the radio, to the infrared.
(Reproduced by permission of AAS.)
        }
\label{fig:sgra_sio}
\end{figure*}

How can one transfer \SgrA's radio position to infrared images with better 
than the 0.01 arcseconds accuracy needed to clearly identify candidates?   
The key is to find sources visible at both radio and infrared wavelengths.
This has been accomplished with red giant stars that are bright
in the infrared and have molecular maser emission at radio wavelengths
from their circumstellar material.  Fig.~\ref{fig:sgra_sio} shows an infrared
image of the central $\pm20$~arcseconds of the Galaxy and the positions
and measured motions of SiO maser emission at radio 
wavelengths for nine stars.  

\begin{figure*}
\begin{center}
\begin{tabular}{cc}
\includegraphics[width=0.90\textwidth]{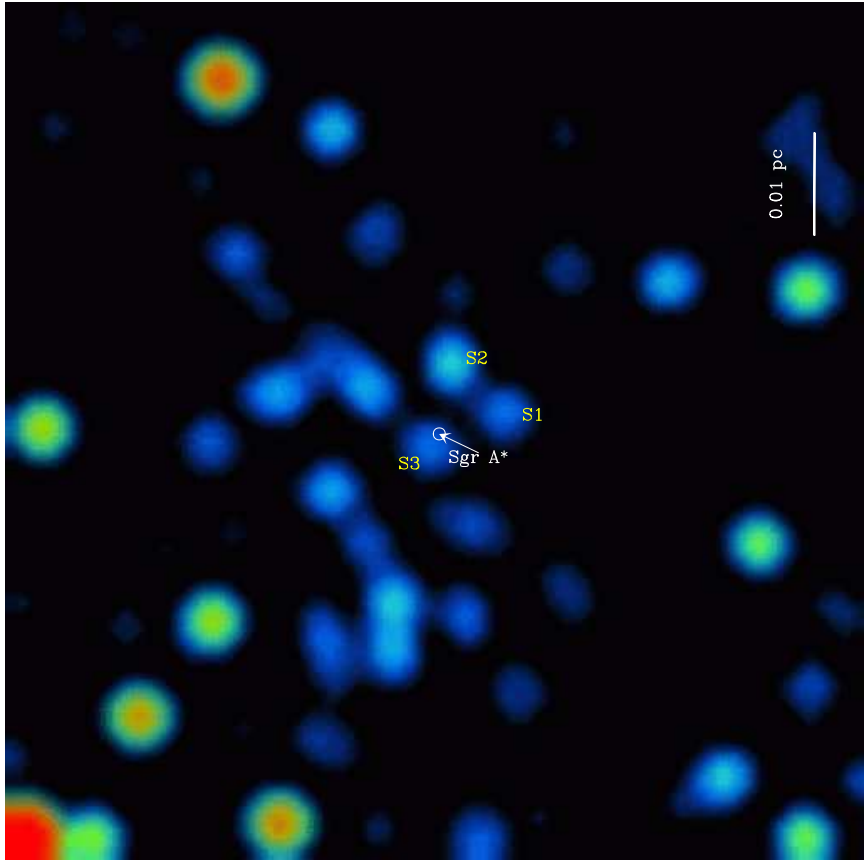}  \\
\end{tabular}
\end{center}
\caption{An infrared frame from July 1995 covering the 
inner $\pm1$ arcsecond from Reid \etal\ (2003).  
Three of the stars that orbit the Galactic center, S1, S2, and S3, are labeled,
as is the position of \SgrA.  The image of star S3 overlaps slightly 
with \SgrA, whose emission is much weaker than a single star.
(Reproduced by permission of AAS.)
        }
\label{fig:sgra_ir}
\end{figure*}

This novel combination of infrared and radio observations has allowed the 
position of the compact radio source at the center of the Galaxy, \SgrA, 
to be transfered accurately ($\pm0.01$~arcseconds or $\pm80$~AU) 
to infrared images\cite{Menten:97,Reid:03,Reid:07}, where \SgrA\ is 
extremely dim.  The position of \SgrA, determined in this manner, 
is circled at the center Fig.~\ref{fig:sgra_sio}.
Fig.~\ref{fig:sgra_ir} shows one 
infrared frame covering the inner-most $\pm1$~arcsecond 
($\pm0.04$~parsecs) of the Galactic center.  Many stars, including some 
that are known to orbit \SgrA, are visible.  However, at the position of 
\SgrA\ no steady emission was detected, although on this frame the 
diffraction-limited image of star S3 overlaps slightly with \SgrA.
Recently, using improved infrared observing techniques, some weak
varying emission has been seen toward the position of \SgrA.
(see \S\ref{section:other}).  
{\it However, the infrared emission of \SgrA\ is much weaker than individual stars!}

The radio position of \SgrA\ transfered to infrared images matches 
the focal position determined from the stellar
orbital fits to within the 0.01~arcseconds measurement accuracy.  
Could \SgrA\  be projected toward the Galactic
center, but in reality not be located there?  
The fact that \SgrA, a nearly unique source in the 
Milky Way, lies in projection within 0.01~arcseconds
(ie, within a solid angle of $\sim10^{-14}$ steradians) of the gravitational 
focus of the stellar orbits makes it extraordinarily unlikely that
it is simply a projection effect.  In addition, \SgrA's apparent motion  
with respect to distant quasars is consistent with it being at the 
dynamical center of the Milky Way (see \S\ref{section:motionless}), 
ensuring that it is indeed at that location.

\subsection  {\SgrA's Emission is Extremely Compact}
\label{section:sgra_size}

The cm-wavelength emission from \SgrA\  is strongly affected by
scattering from interstellar electrons, which increases its apparent size.
At mm-wavelengths the scattering is reduced, and the true size of
the source can be measured.  Many groups have analyzed VLBI data for 
\SgrA\  and find that the intrinsic size is less than 1 AU at mm-wavelengths
\cite{Rogers:94,Krichbaum:98,Doeleman:01,Bower:04,Shen:05}.
However, the absence of strong refractive scintillations for \SgrA\ 
provides a lower limit of $\sim0.1$~AU for the intrinsic size
of the cm-wave emission\cite{Gwinn:91}. 
Thus, the intrinsic radius of the radio emission from \SgrA\  
near 1~cm wavelength is between 1 and 6 Schwarzschild radii 
for a $4\times10^6$~\Msun\ black hole.

\subsection  {\SgrA\ is Motionless at Dynamical Center of Milky Way}
\label{section:motionless}

How much of the unseen mass in the central 100 AU region can be tied
directly to \SgrA?   Were \SgrA\  just a stellar mass object in the 
central stellar cluster, it would be moving at $\sim10^4$ \kms\ 
in the strong gravitational potential as stars are observed to do.  
Only if \SgrA\ is extremely massive could it move slowly.
Early measurements of \SgrA's motion revealed it to be moving $<20$ \kms\ 
\cite{Backer:99,Reid:99}.
  
\begin{figure*}
\begin{center}
\begin{tabular}{cc}
\includegraphics[width=0.9\textwidth]{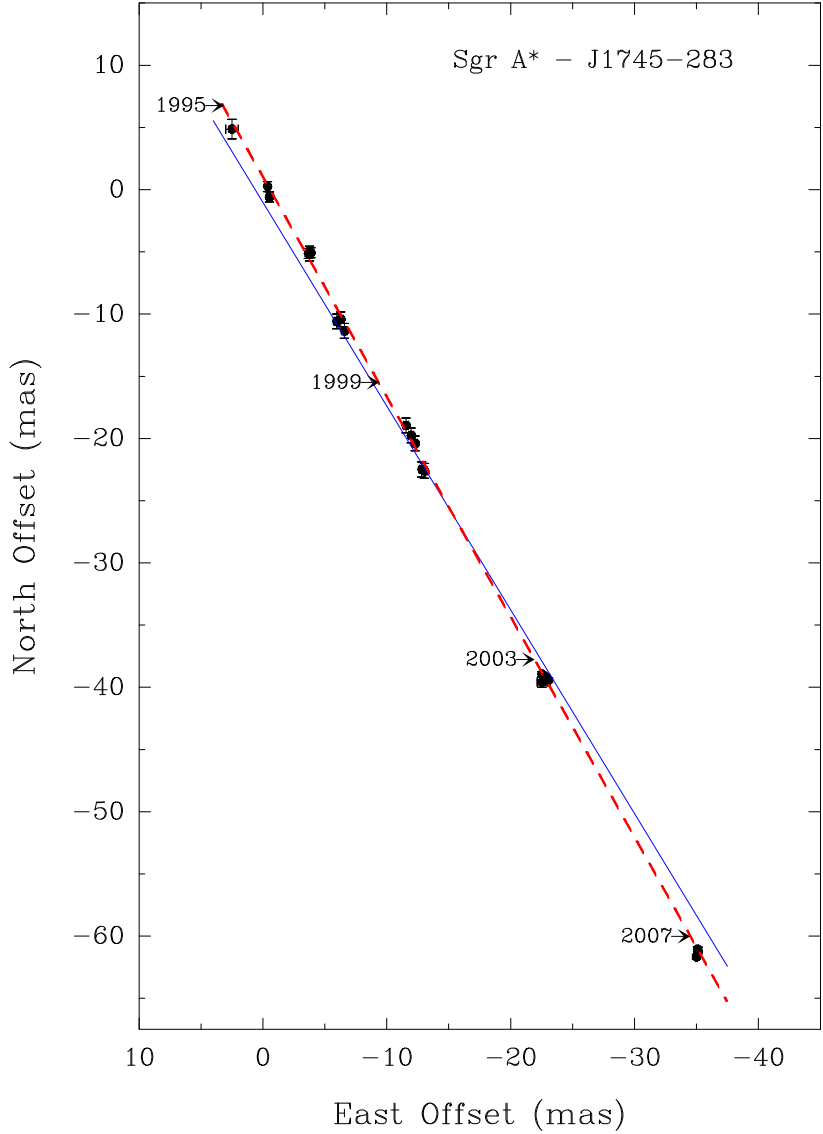}
\\
\end{tabular}
\end{center}
\caption{The {\it apparent} motion on the sky of the compact radio source \SgrA\
relative to a distant quasar (J1745--283).  The dashed line is the
variance-weighted best-fit motion of $0.006379\pm0.000024$
arcseconds per year for the data published by Reid \& Brunthaler through 
2004.  Recent data from 2007 shown here confirms the published result.
All of the {\it apparent} motion of \SgrA\ can be accounted for by the 
$\approx210$ My period orbit of the Sun about the Galactic center.
The solid line gives the orientation of the Galactic plane, and the 
difference in orientation of the two lines is caused by the 7.16 \kms\ 
motion of the Sun perpendicular to the Galactic plane.  The residual,
intrinsic motion of \SgrA\ perpendicular to the Galactic plane is
extremely small: $-0.4\pm0.9$ \kms.  A SMBH perturbed by stars orbiting 
within its gravitation sphere of influence is expected to move $\approx0.2$
\kms\ in each coordinate.  Only a supermassive object could
be motionless in the presence of the $4\times10^6$ \Msun\ known from
infrared stellar orbital fits to occupy this region.
        }
\label{fig:sgra_motion}
\end{figure*}

Fig.~\ref{fig:sgra_motion} shows the latest data on the apparent motion
of \SgrA\ on the sky relative to a background quasar.  This includes
the data published by Reid \& Brunthaler in 2004\cite{Reid:04} and new data 
taken in 2007 that confirms the published results.  The apparent movement 
is as expected for an object that is stationary at the dynamical center of 
the Milky Way and viewed from the Sun-Earth system, which orbits the 
Galaxy with a period of $\approx210$ My at a radius of about 8~kpc and 
a speed of 240~\kms.  It is amazing that VLBA can detect this motion
in only a few weeks time.

The Sun's orbit is almost entirely in the plane of the Galaxy, but  
the orbital speed in the plane is not easily measured from within the 
Galaxy and is uncertain by $\approx20$ \kms.  However, the small component 
of the Sun's motion {\it perpendicular} to the plane of the 
Galaxy is very accurately known ($7.16\pm0.38$~\kms) from observations of 
$10^4$ stars in the solar neighborhood by the astrometric satellite 
HIPPARCOS\cite{Dehnen:98}.  Thus, the contribution of the Sun's motion
perpendicular to the plane of the Galaxy can be removed from the apparent 
motion of \SgrA\  with very high accuracy.  When this is done,
\SgrA's intrinsic motion perpendicular to the plane of the Galaxy is 
$-0.4\pm0.9$ \kms~\cite{Reid:04}.  The extremely small intrinsic motion 
(essentially an upper limit) for \SgrA\  is close to the expected motion 
for a SMBH in the presence of its dense cluster of surrounding stars.

The discovery that \SgrA\  is nearly stationary at the Galactic center
requires that \SgrA\  must contain a significant fraction of the unseen mass 
indicated by the orbiting stars.  The calculation of the expected motion of a 
massive object within its bound stellar cluster (which allows the 
limit on the motion of \SgrA\  to give a minimum mass) 
can be cast as a gravitational Brownian-motion problem
\cite{Chatterjee:02,Dorband:03,Merritt:07}.  
Such a system comes into ``thermal'' equilibrium,
resulting in equipartition of kinetic energy among its constituents.
For the Galactic center case in particular, both an analytical result, 
with only a few simplifications, and a numerical result for fully 
realistic cases allow estimation of the minimum expected motion of a SMBH 
surrounded by a dense stellar cluster\cite{Reid:04}.  
For a $4\times10^6$~\Msun\ black hole, one expects $\approx0.2$~\kms\
for each component of its motion.  This is a very conservative estimate
as it does not include possible clustering of the perturbing stars
nor any contribution from a cluster of dark stellar remnants, which
are expected to have accumulated in this region\cite{Morris:93}.  
Including either of these effects would increase the expected motion of 
the black hole, bringing it very close to the measured limit.

Both analytical and numerical estimates of the mass of the motionless
radio source \SgrA\ yield $>4\times10^5$ \Msun.  
Assuming this mass is contained within 
the observed source size (see \S\ref{section:sgra_size}), 
the mass density is a staggering $>7\times10^{21}$ \Msunpercpc.  
This density is within about three 
orders of magnitude of the ultimate limit of $4\times10^6$ \Msun\ within 
its Schwarzschild radius, $\Rsch$.  For the simplest case of a non-rotating 
black hole, stable gravitational orbits exist only outside of $3\Rsch$.  
Any material inside $3\Rsch$ cannot orbit and falls rapidly into the hole.  
Thus, the effective volume of such a black hole is $3^3$-times that 
defined by the Schwarzschild radius, resulting in a corresponding decrease 
in the ``ultimate''  mass density.  In this case,
the measured density is within only two orders of magnitude of the
black hole limit.  
{\it At this density the evidence is overwhelming that Sgr~A* is a supermassive black hole}
(see \S\ref{section:alternatives}).  

\begin{table}[ph]
\tbl{Density Limits for SMBH Candidates.}
{
\begin{tabular}{@{}lccccc@{}} 
\toprule
Object & Mass    & Radius & $M/R$   & Density   &Reference \\
       & (\Msun) & (AU)   & (kg/m)       & (\Msunpercpc)        \\
\colrule
Globular Cluster&$\p1\times10^6$ &$2\times10^5$&$\p7\times10^{19}$ &$\p2\times10^5\q$ & \\
M~87           &$\p2\times10^9$ &$4\times10^6$ &$\p7\times10^{21}$ &$\p1\times10^{5\q}$ &\cite{Ford:94,Harms:94} \\
NGC~4258       &$\p4\times10^7$ &$2\times10^4$ &$\p3\times10^{22}$ &$\p1\times10^{10}$ &\cite{Miyoshi:95} \\
\SgrA\ (orbits)&$\p4\times10^6$ &$<100$        &$>5\times10^{23}$ &$>8\times10^{15}$   
  &\cite{Schoedel:02,Ghez:03,Schoedel:03,Ghez:05} \\
\SgrA\ (motion)&$>4\times10^5$  &$<0.5$        &$>1\times10^{25}$ &$>7\times10^{21}$   &\cite{Reid:04} \\
SMBH (3\Rsch)  &$\p4\times10^6$ &$0.24$        &$\p2\times10^{26}$ &$\p7\times10^{23}$ &   \\
\botrule
\end{tabular} 
\label{table:densities}
}
\end{table}

For comparison, Table~\ref{table:densities} lists observed densities,
in standard astronomical units, for dense stellar clusters 
(\ie, globular clusters), some SMBH candidates, and a Schwarzschild 
SMBH within its inner-most stable orbit.  Also listed are the mass-to-radius 
($M/R$) ratios, in standard physical units.  For a black hole, 
$M/3\Rsch = c^2/6G$.

The above density calculation assumes that 
the size of the emitting region is equal to or greater than
the size of the mass in \SgrA.  This is true for almost all astrophysical
sources.  Notable exceptions for radio sources are solar flares
and pulsars.  However, these sources 1) are sporadic (either flares
or pulses), 2) are characterized by highly-polarized gyrosynchroton 
emission and 3) have spectra with flux density falling very rapidly with
increasing observing frequency.  \SgrA\ shares none of these 
characteristics; its radio emission is generally slowly varying,
almost unpolarized, and has a rising spectrum.

\section   {Excluding Alternatives to a SMBH} \label{section:alternatives}

Are there alternatives to a SMBH that are consistent with
the extraordinarily high mass density of \SgrA?   
The most obvious possibility is a cluster of dark stars.
As a point of reference, globular clusters are spherical
collections of upwards of $10^6$ stars within a radius of $\sim1$~parsec.
These are long-lived systems, some nearly the age of the Universe, 
and their stellar densities are truly astounding -- imagine placing 
$\sim10^6$ stars between the Sun and the nearest star!  
However, while such a density is extremely high,
it is nearly a factor of $10^{18}$ times less than 
for a SMBH of \SgrA's mass.  (Also, normal, luminous stars are easily 
ruled out by the dearth of infrared emission from the position of 
\SgrA.)

Dense clusters of stars undergo significant gravitational interactions, 
including core-collapse, collisions and evaporation of 
stars\cite{Binney:87}.  Both the evaporation or collisional 
timescales can provide approximate upper limits for the lifetimes of 
such systems\cite{Maoz:98}.  If the cluster members have typical 
masses of $\sim1$~\Msun, then the evaporation timescale for a cluster
that satisfies the Galactic center stellar orbital data would be $<10^6$~years.  
The existence of a cluster with such a short lifetime is extremely 
implausible.   

Evaporation timescales are approximately inversely proportional to 
the typical mass of a cluster member.  Thus, by postulating very low 
member masses, the evaporation timescale can be made arbitrarily long.  
In order to have an evaporation timescale be a reasonable fraction of the 
age of the Galaxy ($\approx13\times10^9$ years), member masses would need 
to be $<0.001$~\Msun.  Thus, one would need $\sim10^{10}$ planets like 
Jupiter to make a long-lived $4\times10^6$~\Msun\ cluster.  
This would seem impossible to arrange.

The argument for excluding dense stellar clusters based on their rapid 
evaporation timescales is based on the implicit assumption of 
an isolated system.  Perhaps a quasi-steady state condition could
occur for which the cluster is fed stars from the outside at a 
rate comparable to the evaporation rate.  A $4\times10^6$~\Msun\ 
cluster that evaporates in $10^6$~years would require the addition
of only a few stars per year to offset evaporation losses.  
This possibility seems unlikely but has yet to be critically analyzed.

Are there other possibilities for extreme concentrations of matter
at the Galactic center?  Hypothetical concentrations of exotic dark
particles have been considered as alternatives to supermassive
black holes in the centers of galaxies.   For example, a ``ball'' 
of heavy Fermions supported by degeneracy pressure has been proposed
\cite{Viollier:93,Tsiklauri:96,Munyaneza:98,Bilic:99}.  
One attractive aspect of this alternative to supermassive black holes 
was that Fermions of rest energy $\sim15$~keV could naturally explain 
the range of masses ($\sim3\times10^6$ to $\sim3\times10^9$~\Msun) 
observed at the centers of galaxies.  However, a Fermion ball cannot achieve 
extraordinarily high central densities and its gravitational potential
is softer than that of a black hole and flattens at the center.  
In 2002, Munyaneza \& Viollier\cite{Munyaneza:02} 
showed that the then existing stellar motion data could only
be fit by a restricted range of parameter space for a supermassive 
black hole, but could be easily fit by Fermion ball models.  
Subsequent observations of stellar orbits\cite{Ghez:03,Schoedel:03,Ghez:05} 
showed that the supermassive black hole predictions were indeed met,
and that the stars moved with speeds greater than allowed by their 
Fermion ball model.

We note that the density limit employed in this section was the one obtained 
only from the orbits of stars.  The additional constraints
from the radio observations provide nearly a factor of $10^6$ more
stringent limit on density, making all of the above arguments
vastly stronger.   Since the density limits are now within only
about two orders of magnitude of that of a SMBH 
(see Table~\ref{table:densities}), any alternatives to a SMBH must
allow similarly high densities.  At such
densities it is difficult to avoid gravitational collapse and
proposed  alternatives, such as ``boson stars''\cite{Torres:00} and 
``magnetospheric eternally collapsing objects,''\cite{Robertson:03} or
others, possibly involving``new physics,'' would be even more fantastic 
than a ``mundane'' SMBH.

Finally, there is now strong evidence for the existence of an
event horizon in both stellar-mass black holes\cite{Garcia:01,McClintock:04}
and for SMBHs.  Many galactic nuclei, in which high mass-densities
indicative of SMBHs are observed, are under-luminous and best
explained with radiatively inefficient accretion flows in which
energy can vanish through the event horizon\cite{Narayan:08}.  
Without an event horizon, energy liberated by the accretion process
cannot vanish.  Indeed, \SgrA\ is such an under-luminous object, 
and evidence from the dearth of infrared emission and an extremely small 
intrinsic size strongly point to an event horizon\cite{Doeleman:08,Broderick:08}.

\section   {Other Evidence for a SMBH} \label{section:other}

    While quiescent emission from \SgrA\ has been difficult to detect
outside of cm to sub-millimeter wavelengths, it does produce 
detectable ``flares'' of short duration at 
radio\cite{Herrnstein:04,Mauerhan:05,Marrone:06} and
infrared\cite{Genzel:03,Ghez:04,Hornstein:07,Do:08} wavelengths and
x-ray energies\cite{Baganoff:01}.  This flaring is thought to be
associated with material occasionally spiraling inward,
converting a significant fraction of its total energy into heat, and
radiating profusely near the inner-most stable orbit around a
black hole.  Of particular interest is the hint of a quasi-periodic 
flaring at infrared wavelengths\cite{Genzel:03,Do:08}, which could arise during 
the final few orbits of material falling into a black hole.
Since, the radius of the inner-most stable orbit depends on black hole
spin and whether the material is orbiting in a prograde or retrograde 
sense\cite{Misner:73}, this may provide an observational 
approach to measure black hole spin\cite{Genzel:03,Shafee:06,McClintock:06}.

    In 1988, J. G. Hills\cite{Hills:88} considered the fate of 
tightly bound binary stars that encounter a supermassive black 
hole at the Galactic center.  If the encounter is close enough, 
he predicted that one of the stars could become bound to the 
black hole, while the other could be ejected at high speeds of
up to 4000~\kms.  Detailed calculations of ejection rates have now
been made\cite{Yu:03}.  Recently, such hyper-velocity stars have been
discovered leaving the Milky Way\cite{Brown:05}, confirming Hills' 
prediction.  While this may be a strong confirmation of the
existence of a supermassive black hole in the Galactic center, 
it is probable that large numbers of stellar-mass black holes have
migrated to the inner 0.1 parsec of the center and hyper-velocity
stars can also be ejected from interactions with these lower mass black
holes\cite{OLeary:08}.

\section   {Broader Implications of SMBHs at Centers of Galaxies}

The highly complementary discoveries at infrared wavelengths (of
stars orbiting an unseen massive object) and at radio wavelengths
(that the stellar orbital focus coincides with \SgrA, that
the size of \SgrA\ is comparable to \Rsch and that \SgrA\ is essentially 
motionless at the dynamical center of the Galaxy) establish with near certainty
that the center of the Milky Way is anchored by a supermassive
black hole.  Of course the Milky Way cannot be unique in the Universe
in having a supermassive black hole at its center, and, indeed,
the Hubble Space Telescope finds evidence for a SMBH at the
center of all galaxies nearby enough so that the telescope can
resolve the gravitational sphere of influence of the SMBH
\cite{Ferrarese:02}.  While
the centers of most, if not all, galaxies contain a SMBH, this is not to
say that all galaxies are identical.  The mass and spin of the black hole 
and the density of stars and gas within its gravitational sphere of 
influence lead to a rich variety of phenomena, which are collectively 
called active galactic nuclei.  \SgrA, being the nearest SMBH, 
serves as the archetypal source for understanding other 
galaxies.  

Astronomers now believe that when supermassive black holes accrete 
matter at a high rate they become extremely luminous and can outshine 
their entire host galaxy of $\sim10^{11}$ stars (ie, the quasar phenomenon).  
Quasars have been found at high redshifts\cite{Fan:01}, 
indicating they existed as early as
1 billion years after the Big Bang, when the Universe was less than
8\% of its current age.  While it is not yet understood how supermassive
black holes form so quickly, they now appear to be an integral part of 
the generation of structure in the early Universe\cite{Haiman:01}.  

Recent observations show a strong correlation between the masses of 
supermassive black holes and the masses and motions of stellar bulges 
which protrude from disk galaxies\cite{Magorrian:98,Ferrarese:00}. 
These observations and large computational simulations of how galaxies
form indicate that supermassive black holes may shape the evolution 
of galaxies\cite{Springel:05,Robertson:06}.  
Thus, supermassive black holes are not only 
fascinating objects from the perspective of fundamental physics, 
they also are exceptionally important from an astrophysical perspective 
for determining the nature of galaxies.  

\section    {Future Observations}

   We know that galaxies interact gravitationally and sometimes
collide and merge.  The merger of two galaxies, each containing
a SMBH, would be followed by orbital decay of the two nuclei,
owing first to dynamical friction and ultimately to gravitational 
radiation (when a tight black hole binary is reached).  The final
merging of two SMBHs produces very strong gravitational waves,
which might be detected with the future space-based gravity-wave 
detector LISA\cite{LISA:08}.

   Returning to the nearest SMBH, \SgrA, an obvious goal for astronomers 
is to image the region within the inner-most stable orbit with 
resolution of $\sim\Rsch$.
As this review was being assembled, interferometer fringes were reported
toward \SgrA\ at a wavelength of 1.3~mm and a fringe
spacing of $0.00005$~arcseconds\cite{Doeleman:08}.  This demonstrates that
a significant fraction of the mm-wave emission of \SgrA\ comes from
a region with a radius $<0.2$~AU, which is $<3\Rsch$.  When ``snapshot''
images can be made at slightly shorter wavelengths (eg, 0.8~mm),
they should reveal the ``shadow'' caused by extreme bending of
space around a black hole\cite{Falcke:00} and might show highly dynamic 
activity associated with the final in-spiral of plasma toward the
event horizon\cite{Broderick:06}.  

   Astrometric observations of the region within $\sim10\Rsch$ can
also be made with infrared interferometry.  Plans for such an interferometer 
experiment, called GRAVITY\cite{Eisenhauer:08}, are well underway.  
Since the radius, and hence period, of the inner-most stable orbit 
depends on the black hole spin, measuring the in-spiraling material
motion can yield determinations of spin.

   Stars orbiting with short periods ($\sim1$ yr) in highly elliptical 
orbits would approach very close to \SgrA\ at pericenter and should display 
general relativistic precession of their orbital planes.  Diffraction-limited 
imaging of the Galactic center in the infrared with future extremely large 
($>30$~m diameter) telescopes should be able to make such 
observations\cite{Weinberg:05} and also allow a test of the ``no hair''
theorem for a black hole\cite{Will:08}

\section {Summary}

The major observational results that provide overwhelming evidence
that \SgrA\ is a SMBH are as follows:

\begin{itemlist}[(ii)]
\item Stars near \SgrA\ move on elliptical orbits with a common
      focal position.      
\item The required central mass is $4\times10^6$~\Msun\ within a
      radius of 100~AU.
\item The position of \SgrA\ agrees with the orbital focus to within
      measurement uncertainty of $\pm80$~AU.
\item The infrared emission from \SgrA\ is far less luminous
      than a single star.  
\item The intrinsic size of \SgrA\ at mm-wavelengths is $<6\Rsch$.
\item \SgrA\ is intrinsically motionless at the \kms\ level at the
      dynamical center of the Galaxy. 
\end{itemlist}

The great impact of these discoveries is their simplicity and elegance.
Elliptical orbits for stars provide an absolutely clear and unequivocal
proof of a great unseen mass concentration.  The discoveries that the
compact radio source is at the position of the unseen mass and is 
motionless provide even more compelling evidence for a supermassive black hole.
{\it Together they form a simple, unique demonstration that the fantastic 
concept of a supermassive black hole is, with a high degree of certainty,
a reality.}

\section*{Acknowledgements}
I thank R. Genzel and A. Ghez for comments and suggestions on a draft 
of this review.

\end{document}